\documentclass[reprint,superscriptaddress,nofootinbib,aps,prd,10pt]{revtex4-1}
\usepackage{amsmath,amssymb,marvosym}
\usepackage{graphicx}
\usepackage{amssymb}
\usepackage{bbold}
\usepackage{color}
\usepackage{tikz}
\usetikzlibrary{plotmarks}
\newcommand\marksymbol[2]{\tikz[#2,scale=1.2]\pgfuseplotmark{#1};}
\def\spose#1{\hbox to 0pt{#1\hss}}
\def\ltapprox{\mathrel{\spose{\lower 3pt\hbox{$\mathchar"218$}}
 \raise 2.0pt\hbox{$\mathchar"13C$}}}
\def\gtapprox{\mathrel{\spose{\lower 3pt\hbox{$\mathchar"218$}}
 \raise 2.0pt\hbox{$\mathchar"13E$}}}

\newcommand{\1}{\mathbb{1}}
\DeclareMathOperator{\Tr}{Tr}

\begin{document}


\title{{\Large {\bf 
       Bloch Waves in Minimal Landau Gauge and \\[2mm]
       the Infinite-Volume Limit of Lattice Gauge Theory}}}

\author{Attilio~Cucchieri}
\affiliation{Instituto de F\'\i sica de S\~ao Carlos, Universidade de S\~ao Paulo, \\
             Caixa Postal 369, 13560-970 S\~ao Carlos, SP, Brazil}
\author{Tereza~Mendes}
\affiliation{Instituto de F\'\i sica de S\~ao Carlos, Universidade de S\~ao Paulo, \\
             Caixa Postal 369, 13560-970 S\~ao Carlos, SP, Brazil}

\begin{abstract}
By exploiting the similarity between Bloch's theorem for electrons in 
crystalline solids and the problem of Landau gauge-fixing in Yang-Mills
theory on a ``replicated'' lattice, one is able to obtain essentially
infinite-volume results from numerical simulations performed on a 
relatively small lattice. This approach,
proposed by D.\ Zwanziger in \cite{Zwanziger:1993dh},
corresponds to taking the infinite-volume limit for Landau-gauge field
configurations in two steps: firstly for the gauge transformation alone,
while keeping the lattice volume finite, and secondly for the gauge-field 
configuration itself. The solutions to the gauge-fixing 
condition are then given in terms of Bloch waves.
Applying the method to data from Monte Carlo 
simulations of pure SU(2) gauge theory in two and three space-time 
dimensions, we are able to evaluate the Landau-gauge gluon
propagator for lattices of linear extent up to sixteen times
larger than that of the simulated lattice.
The approach is reminiscent of Fisher and Ruelle's construction of the 
thermodynamic limit in classical statistical mechanics.
\end{abstract}

\maketitle


\section{Introduction}

Since 2007 \cite{2007,Cucchieri:2007md}, we know that 
very large (physical) volumes are required in
lattice simulations of Yang-Mills theories in minimal Landau gauge 
if one wishes to uncover the true infrared behavior
of Green's functions, a topic that has attracted much attention 
in the past two decades \cite{reviews}. 
Indeed, due to the generation of a dynamical mass $m_g$ of a few hundred MeV 
in the gluon sector (see \cite{Cucchieri:2003di,gluonmass,Cucchieri:2011ig}
and references therein), one must reach
momenta $p$ as small as 50 MeV
in order to generate data at $p \ll m_g$,
allowing a good description of gluonic correlation 
functions in the infrared limit. 
Since the smallest momentum 
on a lattice with $N$ points per direction is $\,2 a^{-1}\sin{(\pi / N)}$,
where $a$ is the lattice spacing, the above requisite implies an 
unfeasibly large lattice side of about 250 points
(for $a\approx 0.1$ fm).
Alternatively, one needs good control of the extrapolation
of the data to infinite volume. A step in this direction
was the consideration of exact (upper and lower) bounds for
gluon and ghost propagators \cite{Cucchieri:2007rg,bounds}, which can help 
in extrapolating the numerical data to large lattice volumes.

In this work, following an idea of D.\ Zwanziger
\cite{Zwanziger:1993dh}, we present a new approach,
based on taking the infinite-volume limit for a
Landau-gauge transformation applied to a (replicated) 
thermalized field configuration at a given volume $V$.
The corresponding setup allows one to prove a 
result similar to Bloch's theorem for crystalline solids. 
As a consequence, even though one deals with gauge 
transformations on the extended lattice, the numerical gauge
fixing is actually done on the original (small) lattice.
The obtained gauge transformation is then used to evaluate 
a Landau-gauge-fixed gluon-field configuration and the corresponding 
gluon propagator in momentum space $D(p^2)$.

The paper is organized as follows.
In the next section, we briefly review Bloch's theorem and
discuss the idea presented in \cite{Zwanziger:1993dh}. 
We then show our preliminary numerical results and,
in the last section, we draw our conclusions.


\section{Bloch waves}

An ideal crystalline solid in $d$ dimensions is (geometrically) defined 
by a Bravais lattice \cite{AM}:
an infinite set of points $\vec{R} = n_{\mu} \vec{a}_{\mu} $, 
where $n_{\mu} \in \cal{Z}$ (with $\mu = 1, \ldots, d$),
$\vec{a}_{\mu}$ are $d$ linearly independent vectors, 
and the sum over repeated indices is understood. 
For Bloch's theorem one also
considers an electrostatic potential $U(\vec{r})$, due to the
ions of the solid, with the periodicity of the Bravais lattice,
i.e.\ $U(\vec{r}) = U(\vec{r} + \vec{R})$ for any Bravais-lattice 
vector $\vec{R}$. The corresponding Hamiltonian
$\cal{H}$ for a single electron is then invariant under translations 
by $\vec{R}$ ---represented by the operators ${\cal T}(\vec{R})$---
and we can choose the eigenstates $\psi(\vec{r})$ of $\cal{H}$ 
to be also eigenstates of ${\cal T}(\vec{R})$. Now, since
\begin{equation}
{\cal T}(\vec{R}) \, {\cal T}(\vec{R}^{'}) \, = \,
{\cal T}(\vec{R}^{'}) \, {\cal T}(\vec{R}) \, = \,
{\cal T}(\vec{R} + \vec{R}^{'}) \; ,
\label{eq:TT}
\end{equation}
${\cal T}(\vec{R})$ has eigenvalues
$\exp{( i \vec{k} \cdot \vec{R} )} =
\exp{\left( 2 \pi i \, k_{\nu} \, n_{\nu} \right)}$, i.e.\ 
${\cal T}(\vec{R}) \,
\psi(\vec{r}) = \psi(\vec{r} +\vec{R}) = \exp{( i \vec{k}
\cdot \vec{R} )} \, \psi(\vec{r})$. Here
$\vec{k} = k_{\nu} \vec{b}_{\nu}$ is a vector of the
reciprocal lattice: $k_{\nu} \in {\cal Z}$ (with $\nu
= 1, \ldots, d$) and $\vec{a}_{\mu} \cdot \vec{b}_{\nu}
= 2 \pi \delta_{\mu \nu}$, usually restricted to the first
Brillouin zone. 
As a consequence,
the eigenstates $\psi(\vec{r})$ can be written as Bloch waves
\begin{equation}
\psi_{\vec{k}}(\vec{r}) \, =\, \exp{( i \vec{k} \cdot \vec{r} )}
              \, h_{\vec{k}}(\vec{r}) \; ,
\label{eq:bloch}
\end{equation}
where the functions $h_{\vec{k}}(\vec{r})$ have the periodicity
of the Bravais lattice, i.e.\ $h_{\vec{k}}(\vec{r} + \vec{R}) =
h_{\vec{k}}(\vec{r})$.

Let us now consider a thermalized link configuration $\{
U_{\mu}(\vec{x}) \}$, for the SU($N_c$) gauge group in
$d$ dimensions, defined on a lattice $\Lambda_x$
with volume $V=N^d$ and periodic boundary conditions (PBC).
Then, following Ref.\ \cite{Zwanziger:1993dh}, we extend
$\Lambda_x$ by replicating it $m$ times along each
direction, yielding an extended lattice $\Lambda_z$,
with lattice volume $m^d V$ and PBC.
Let us note that a similar idea has been recently used
in Ref.\ \cite{Lehner:2015bga} in order to include
infinite-volume QED effects into a finite QCD system.
We indicate the points of $\Lambda_z$ with
\begin{equation}
\vec{z} \, = \, \vec{x} \, + \, \vec{y} N \; ,
\label{eq:zcoord}
\end{equation}
where $\vec{x}\in \Lambda_x$ and
$\vec{y}$ belongs to the {\em replica lattice} $\Lambda_y$:
$y_{\mu} = 0, \ldots, m-1$.
By construction, $\{ U_{\mu}(\vec{z}) \}$ 
is invariant under translation by $N$ in any direction.

We now impose the minimal-Landau-gauge condition on $\Lambda_z$, i.e.\
we consider the minimizing functional
\begin{equation}
{\cal E}_U[g] \, = \, - \frac{\Re \, \Tr}{d\,N_c\,m^d V} \,
\sum_{\mu = 1}^d \, \sum_{\vec{z}\in\Lambda_z}
\, g(\vec{z})
\, U_{\mu}(\vec{z}) \, g(\vec{z} + \hat{e}_{\mu})^{\dagger}
\; ,
\label{eq:minimizing}
\end{equation}
where $g(\vec{z})$ are SU($N_c$) matrices,
$\hat{e}_{\mu}$ is a unit vector in the $\mu$ direction, $\Re
\,\Tr$ indicates the real part of the trace and $^{\dagger}$
stands for the Hermitian conjugate. Also, the
minimization is done with respect to the gauge transformation
$\{ g(\vec{z}) \}$, with the link configuration $\{
U_{\mu}(\vec{z}) \}$ kept fixed. The resulting gauge-fixed
field configuration is transverse on $\Lambda_z$.
Note that for $\{ g(\vec{z}) \}$ we take PBC on $\Lambda_z$,
i.e.\ $g(\vec{z}) = g(\vec{z} + m N \hat{e}_{\mu})$ for 
$\mu = 1, \ldots, d$.

The analogy of the above minimization problem with the
setup for Bloch's theorem is evident: $\Lambda_y$ is a finite 
Bravais lattice with PBC and the thermalized lattice
configuration $\{ U_{\mu}(\vec{z}) \}$ corresponds to the
periodic electrostatic potential $U(\vec{r})$. It is then not
surprising that one can prove \cite{Zwanziger:1993dh}, in
analogy with Eq.\ (\ref{eq:bloch}), that the gauge
transformation $g(\vec{z})$ that yields a given local
minimum of ${\cal E}_U[g]$ can be written as
\begin{equation}
g(\vec{z}) \, =\, e^{i \Theta_{\mu} \, z_{\mu} / N}
              \, h(\vec{z})
\, =\, e^{i \Theta_{\mu} \, z_{\mu} / N}
              \, h(\vec{x})  \; ,
\label{eq:bloch-g}
\end{equation}
where we make explicit that $h(\vec{z}) \in$ SU($N_c$)
is invariant under a shift by $N$, i.e.\
$h(\vec{z} + N \hat{e}_{\mu}) = h(\vec{z})$. Here, the vectors
$\vec{z}$ and $\vec{x}$ are related through Eq.\ (\ref{eq:zcoord}) and
the matrices $\Theta_{\mu}$ ---having eigenvalues $2 \pi n_{\mu} / m$
(with $n_{\mu} \in {\cal Z}$)--- can be written as $\tau^a
\theta^a_{\mu}$ (with $a=1,\ldots,N_c-1$), where the
$\tau^a$ belong to a Cartan sub-algebra of the SU($N_c$) Lie
algebra. It is important to note that, due to Eq.\
(\ref{eq:bloch-g}) and to cyclicity of the trace, the
minimizing functional ${\cal E}_U[g]$ in Eq.\ (\ref{eq:minimizing}) becomes
\begin{eqnarray}
{\cal E}_U[g] & = & - \frac{\Re \, \Tr}{d\,N_c\,V} \,
 \sum_{\mu = 1}^d \, e^{- i \Theta_{\mu} / N} \,
  Q_{\mu}  \; ,
\label{eq:minimizing2} \\[2mm]
  Q_{\mu} & = & \sum_{\vec{x}\in\Lambda_x}
   \, h(\vec{x}) \, U_{\mu}(\vec{x}) \,
h(\vec{x} + \hat{e}_{\mu})^{\dagger} \; ,
\label{eq:Q}
\end{eqnarray}
i.e.\ the numerical minimization, which now includes
extended gauge transformations, can still be carried out
on the original lattice $\Lambda_x$.

The proof of Eq.\ (\ref{eq:bloch-g}) is quite similar---see
Appendix F of Ref.\ \cite{Zwanziger:1993dh}---to the proof
of Bloch's theorem. Indeed, the minimizing problem
(\ref{eq:minimizing}) is clearly invariant if we consider a
shift of the lattice sites $\vec{z}$ by $N$ in any direction
$\hat{e}_{\mu}$, since this amounts to a simple redefinition
of the origin for $\Lambda_z$. Note also that,
due to cyclicity of the trace, 
${\cal E}_U[g]$ is invariant under (left) global
transformations and thus 
$\{ g(\vec{z})\}$ is determined modulo a global transformation.
As a result, if $\{ g(\vec{z}) \}$ is unique (see discussion below),
we must have
\begin{equation}
{\cal T}(N \hat{e}_{\mu}) \, g(\vec{z}) \, = \,
 g(\vec{z} + N \hat{e}_{\mu}) \, = \,
\lambda_{\mu} \, g(\vec{z}) \; ,
\label{eq:vmu}
\end{equation}
where $\lambda_{\mu}$ is a $\vec{z}$-independent 
SU($N_c$) matrix.
At the same time, by using the relation
(\ref{eq:TT}) for the translation operators, we obtain that the
$\lambda_{\mu}$'s are commuting matrices, i.e.\ they can be
written as $\exp{( i \Theta_{\mu} )} = \exp{( i \tau^a
\theta^a_{\mu} )}$, where the $\tau^a$ matrices are Cartan
generators. Then, by using Eq.\ (\ref{eq:zcoord}) and applying 
Eq.\ (\ref{eq:vmu}) reiteratively, we find
\begin{equation}
g(\vec{z}) \, = \, \exp{( i \Theta_{\mu} y_{\mu} )} \,
g(\vec{x}) \; ,
\label{eq:gofz}
\end{equation}
where the gauge transformation $g(\vec{x})$ is
defined on the first lattice $\Lambda_x$ of $\Lambda_z$ (corresponding
to $y_{\mu} = 0$ for all directions $\mu$). Thus, Eq.\
(\ref{eq:bloch-g}) is immediately obtained if one writes
\begin{equation}
g(\vec{x})  \,\equiv\, \exp{( i \Theta_{\mu}
               x_{\mu} / N )} \, h(\vec{x}) \; ,
\label{eq:gofx}
\end{equation}
yielding Eqs.\ (\ref{eq:minimizing2}) and (\ref{eq:Q}).
Moreover, due to the PBC for $\Lambda_z$, we need to
impose the conditions $\left[ \, \exp{( i \Theta_{\mu} )}
\,\right]^m = \1$, where $\1$ is the identity matrix.
Clearly, these conditions are satisfied if the
eigenvalues of the matrices $\Theta_{\mu}$
are of the type $2 \pi n_{\mu} / m$, with $n_{\mu} \in
{\cal Z}$.

In the SU(2) case, considered here, a Cartan sub-algebra
is one-dimensional and, by taking the third Pauli matrix
$\sigma_3$ as the Cartan generator, one can write
\cite{Zwanziger:1993dh} the most general gauge
transformation (\ref{eq:bloch-g}) by considering
$\Theta_{\mu} = 2 \pi ( v^{\dagger} \sigma_3 v ) \,
n_{\mu} / m$ with $v \in $ SU(2).

Before presenting the numerical results obtained with
the new approach described above, let us discuss the
hypothesis of uniqueness for the gauge transformation
$\{ g(\vec{z}) \}$, which is essential for Eq.\ (\ref{eq:vmu})
to be valid. In Ref.\ \cite{Zwanziger:1993dh} the gauge
fixing on $\Lambda_z$ is considered only for the absolute
minima of the minimizing functional, belonging to the interior
of the so-called fundamental modular region. Since these
minima are proven to be unique (see Appendix A of the same
reference), the implicit assumption made in \cite{Zwanziger:1993dh}
is that the gauge transformation $\{ g(\vec{z}) \}$ that connects
the unfixed, thermalized configuration $\{ U_{\mu}(\vec{z}) \}$ to
the (gauge-fixed) absolute minimum $\{U^{(g)}_{\mu}(\vec{z}) \}$
is also unique, modulo a global transformation, thus
implying Eq.\ (\ref{eq:vmu}). However,
the same hypothesis also applies to a
specific local minimum. Indeed, even though local minima
can be degenerate, a specific realization of one of these
minima also requires a specific and unique
$\{ g(\vec{z}) \}$ (up to a global transformation)
when starting from a given $\{ U_{\mu}(\vec{z}) \}$.


\begin{figure}
\centering
\hskip -2.5mm
\includegraphics[trim=25 0 40 0, clip, scale=0.85]{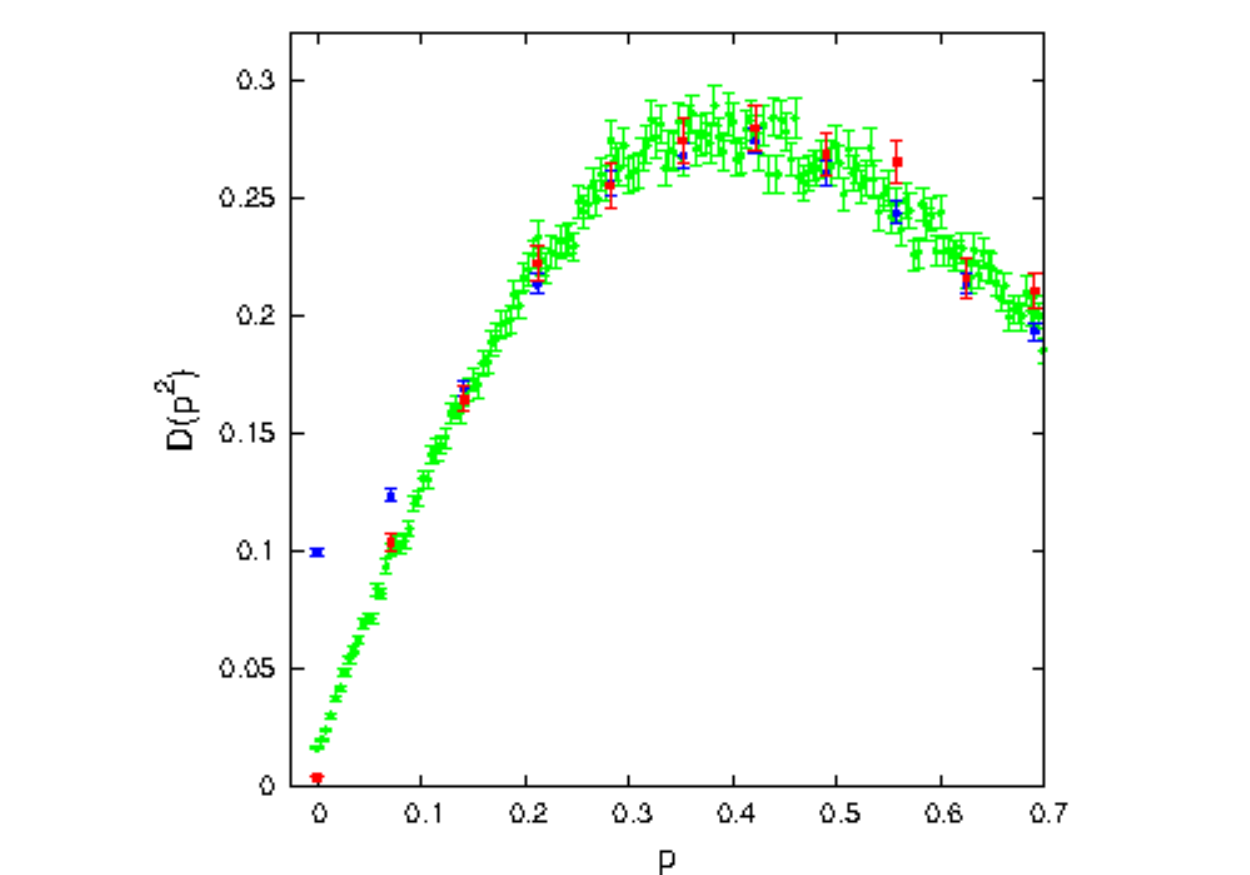}
\vskip 3mm
\includegraphics[trim=25 0 40 0, clip, scale=0.85]{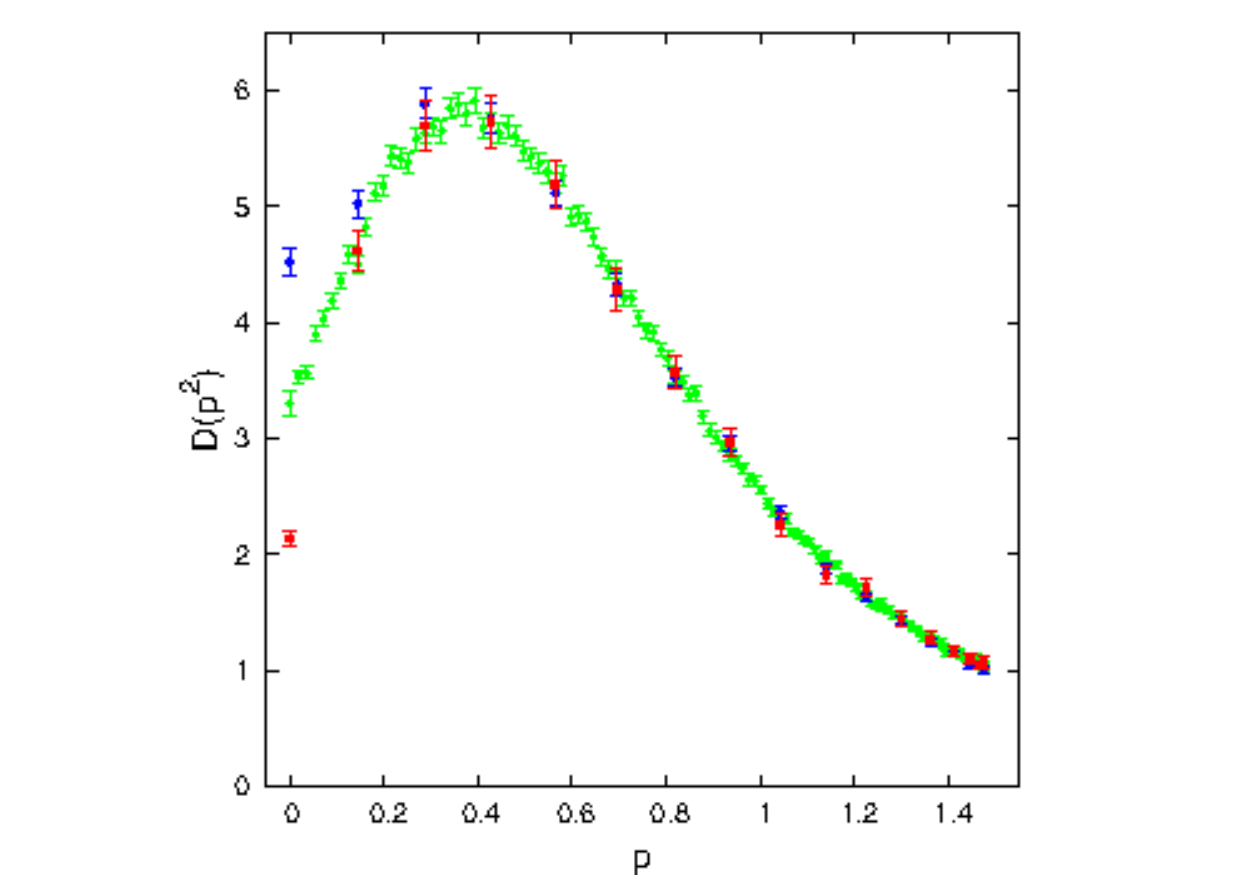}
\caption{\label{fig} The gluon propagator $D(p^2)$ as a function of
    the lattice momentum $p$. {\bf Top}:
    $d=2$ case with $\beta = 10.0$, considering original 
    $\Lambda_x$ lattice volumes $80^2$ (\protect\marksymbol{oplus*}{blue})
    and $1280^2$ (\protect\marksymbol{diamond*}{green}), and an extended
    $\Lambda_z$ lattice volume $80^2 \times 16^2 = 1280^2$
    (\protect\marksymbol{square*}{red}). {\bf Bottom}: 
    $d=3$ case with $\beta = 3.0$, considering original
    $\Lambda_x$ lattice volumes $32^3$ (\protect\marksymbol{oplus*}{blue})
    and $256^3$ (\protect\marksymbol{diamond*}{green}) and an extended
    $\Lambda_z$ lattice volume $32^3 \times 8^3 = 256^3$
    (\protect\marksymbol{square*}{red}). 
}
\end{figure}

\section{Numerical simulations}
\label{sec:data}

From the practical point of view, the minimization
of the functional ${\cal E}_U[g]$, defined in Eqs.\
(\ref{eq:minimizing2}) and (\ref{eq:Q}), can be
done recursively, using two alternating steps:
{\bf a)} the matrices $\Theta_{\mu}$
are kept fixed as one updates the matrices
$h(\vec{x})$ by sweeping through the lattice
using a standard gauge-fixing algorithm
\cite{gaugefixing}
and {\bf b)} the matrices $Q_{\mu}$ are
kept fixed as one minimizes ${\cal E}_U[g]$
with respect to the matrices $\Theta_{\mu}$,
belonging to the corresponding Cartan sub-algebra.
After the gauge fixing is completed, one can evaluate
the gauge-transformed link variables $U^{(g)}_{\mu}(\vec{z}) =
g(\vec{z}) \, U_{\mu}(\vec{z}) \,
g(\vec{z} + \hat{e}_{\mu})^{\dagger}$. Then, considering
Eqs.\ (\ref{eq:gofz}) and (\ref{eq:gofx}), and the invariance
of the link configuration $\{ U_{\mu}(\vec{z}) \}$ under
translation by $N$,
it is clear that
the dependence of
$U^{(g)}_{\mu}(\vec{z})$ on the $y_{\mu}$
coordinates is rather trivial. As a consequence,
the gluon
propagator evaluated with extended gauge transformations
is nonzero only for a subset of the lattice momenta
available on the extended $\Lambda_z$ lattice
\cite{prep}.

Here we present data for the two- and the three-dimensional
cases, for which it is feasible to simulate
at considerably large lattice volumes (without the use
of extended gauge transformations). This allows a comparison
of the new approach with the traditional method at
small momenta, for which finite-size effects are larger.
Indeed, such effects are strongest in the $d=2$ case, since the gluon 
propagator is
of the scaling type \cite{Cucchieri:2007rg,2d,Cucchieri:2011ig},
i.e.\ $D(0)=0$ in the infinite-volume limit. The
effects are also very large in the $d=3$ case, for which the gluon
propagator is of the massive type \cite{Cucchieri:2007md,
Cucchieri:2011ig} but with a clear and pronounced turnover
point at small momenta \cite{Cucchieri:2003di,Cucchieri:2011ig}.

As one can see in Fig.\ \ref{fig},
the results obtained for an extended lattice $\Lambda_z$
show very good agreement with the ones obtained 
with the traditional method for the same lattice volume, while
the results from the corresponding original lattice $\Lambda_x$
deviate considerably from the large-lattice results in the infrared limit.
We note that the strong suppression of $D(0)$
is a peculiar effect of the extended gauge transformations
\cite{prep}.


\section{Conclusions}

We have investigated an analogue of Bloch's theorem for (lattice) 
Landau gauge-fixing \cite{Zwanziger:1993dh}, which
arises because the Landau-gauge condition leaves a residual global
transformation unfixed.
We find that, at least in the gluon sector, numerical
results for large lattice volumes can be reproduced
by simulations at much smaller volumes with extended gauge 
transformations,
thus reducing memory usage by a huge factor (at least up to $16^d$
for $d=2,3$). The only limitation of the approach in its present form
is that the allowed momenta are set by the discretization
on the original (small) lattice $\Lambda_x$
(see Fig.\ \ref{fig} and discussion in the previous section).

As observed for the limiting sequence of domains in the construction
of infinite-volume statistical mechanics \cite{Ruelle}, 
our results show that the information encoded in a thermalized 
gauge-field configuration does not depend much on the (original)
lattice volume $V$ considered.
As a consequence, the properties of Landau-gauge Green's functions
are essentially determined by the gauge-fixing procedure and, in this case,
the size of the (extended) lattice volume matters! 
Let us note that another illustration of the nontrivial r\^ole of gauge-fixing 
is the fact that the lattice Landau-gauge gluon propagator at 
$\beta =0$, i.e.\ for a completely random link configuration 
$\{ U_{\mu}(\vec{x}) \}$, shows qualitative agreement with the
one at nonzero $\beta$ \cite{Cucchieri:2009zt}.

More results and details of the numerical simulations presented
here will be discussed elsewhere \cite{prep}. We also plan to extend 
our study of Green's functions using extended gauge transformations
to the ghost sector, investigating the impact on properties of the first 
Gribov region, and to the matter sector. A possible improvement of the
approach is the use of continuous external momenta \cite{externalmom},
which could make the method more attractive in the $d=4$ case. 


\vspace{6mm}
\section*{Acknowledgments}
We thank B. Blossier for discussions and CNPq for partial support.
Simulations were done on the IBM cluster at USP
(FAPESP grant \# 04/08928-3) and on the
Blue Gene/P supercomputer supported by the Research Computing Support 
Group (Rice University) and Laborat\'orio de
Computa\c c\~ao Cient\'ifica Avan\c cada (Universidade de S\~ao Paulo).



\end{document}